\pdfoutput=1
\documentclass[runningheads]{llncs}

\usepackage[T1]{fontenc}
\usepackage{graphicx}
\usepackage{booktabs}
\usepackage{array}
\usepackage{microtype}

\usepackage{xurl}
\usepackage[hidelinks]{hyperref}
\begin{document}

\title{Human-Centred Risk Mitigation for AI-Mediated Information Manipulation:
A SOCMINT Framework Based on Information Manipulation Sets}
\titlerunning{Human-Centred Risk Mitigation for AI-Mediated Information Manipulation}

\author{Antonio Scala}
\authorrunning{A. Scala}

\institute{Consiglio Nazionale delle Ricerche, Istituto dei Sistemi Complessi, Rome, Italy\\
\email{antonio.scala@cnr.it}}
\maketitle
\begin{abstract}
AI-mediated information manipulation increasingly takes the form of social cyber attacks that target trust, attention, credibility, reputation, and decision-making rather than only technical infrastructures or isolated false contents. Existing defensive approaches often oscillate between incident-level analysis, which fragments campaigns into weak signals, and attribution-first analysis, which may delay mitigation until responsibility is established. This paper proposes a SOCMINT framework based on Information Manipulation Sets (IMS) as an intermediate operational unit between individual incidents and strategic attribution. Building on the VIGINUM/EEAS use of IMS in counter-FIMI analysis, the framework treats manipulation as a coherent process involving narratives, accounts, infrastructures, temporal patterns, cross-platform migration, synthetic amplification, and cognitive targeting. The proposed pipeline moves from signal detection and diagnostic triage to IMS hypothesis construction, confidence/severity assessment, mitigation selection, and iterative update. A compact scenario illustrates how IMS-based analysis captures what content-level and attribution-first approaches miss. The paper also proposes a tabletop evaluation protocol to assess decision quality, confidence calibration, and mitigation proportionality. The main implication is that human-centred risk mitigation requires not only better detection, but also structured reasoning under uncertainty, auditable decision-making, and safeguards against over-securitising legitimate dissent.
\end{abstract}

\section{Introduction}

Artificial intelligence is increasingly embedded in the production, distribution, filtering, and interpretation of information. Large language models, recommender systems, automated content generation, synthetic personas, and data-driven targeting tools alter the scale, speed, variability, and adaptability of manipulative operations \cite{goldstein2023generative,bontcheva2023generativeAI}.

At the same time, cyber threats are no longer limited to the compromise of technical infrastructures. A growing class of adversarial activities targets trust, attention, credibility, reputation, and decision-making. Phishing, social engineering, disinformation campaigns, coordinated harassment, reputational attacks, and narrative manipulation exploit human behaviour and online communication patterns \cite{nist2023humanCentered,eeas2026fourthFIMI}.

This paper refers to these phenomena as \textbf{AI-mediated social cyber attacks}. The expression identifies adversarial processes in which digital infrastructures, platform dynamics, social relations, and cognitive vulnerabilities are combined to manipulate interpretation and behaviour. The relevant object of analysis is often not an isolated false content, but a process involving accounts, channels, messages, narratives, infrastructures, timing, amplification, and target communities.

This creates an analytical gap. Incident-level analysis can identify false claims, harmful contents, or suspicious accounts, but it often misses the operational structure that connects heterogeneous traces into a manipulation process. Attribution-first analysis may require evidentiary thresholds that are unavailable while the operation is still shaping perception. Between the individual incident and the fully attributed campaign, analysts need an intermediate operational unit.

This paper uses \textbf{Information Manipulation Sets (IMS)} as that unit. It builds on the operational use of IMS in counter-FIMI analysis and repositions it within a SOCMINT framework for AI-enabled social cyber attacks. An IMS groups observable behaviours, narratives, assets, infrastructures, temporal patterns, and amplification mechanisms that appear to belong to a coherent manipulation process. It does not replace attribution; it supports graduated attribution and risk mitigation before the ultimate sponsor is known.

The contribution is not a new machine-learning detector. It is a conceptual-operational framework connecting AI-enabled manipulation, SOCMINT observables, IMS construction, human-in-the-loop risk assessment, and proportionate mitigation. The paper develops an IMS-based SOCMINT framework by positioning IMS between incident-level detection and strategic attribution, defining a risk-mitigation pipeline, organising relevant observables, outlining a tabletop evaluation protocol, and incorporating democratic safeguards into the design.

The paper is organised as follows. Section 2 situates AI-enabled social cyber attacks within the transition from disinformation to socio-technical risk. Section 3 introduces IMS. Section 4 examines AI as an accelerator of manipulation. Sections 5 and 6 present the pipeline and indicators. Section 7 gives an illustrative scenario. Section 8 discusses human-centred safeguards. Section 9 proposes a tabletop evaluation strategy. Sections 10 and 11 discuss limitations and conclude.

\section{Background: From Disinformation to Social Cyber Attacks}

\subsection{Beyond false content}

Disinformation is often approached as a problem of false or misleading content. This perspective is necessary but insufficient. Many manipulative operations do not rely on entirely fabricated claims. They may use authentic materials taken out of context, selective presentation of facts, emotionally charged framing, misleading amplification, or the strategic erosion of trust in reliable sources.

The relevant security question is therefore not only whether a content item is false. It is whether a communication process manipulates the informational conditions under which audiences assign credibility and relevance. A true document can be weaponised through selective disclosure. A legitimate controversy can be amplified until it appears to indicate institutional collapse. A minor error can be turned into a reputational attack. A real crisis can be framed in ways that intensify fear, anger, distrust, or polarisation.

\subsection{From disinformation to social cyber attacks}

The notion of \textbf{social cyber attack} captures a broader adversarial space. It includes operations that exploit digital systems in order to manipulate human and social processes. Phishing and social engineering target trust and attention at the individual or organisational level. Disinformation campaigns target collective interpretation. Coordinated influence operations target public debate, institutional legitimacy, or group behaviour. Reputational attacks target the credibility of persons, organisations, companies, or public authorities.

These operations are cyber-related because they depend on digital infrastructures, platforms, data flows, automation, and networked communication. Yet their effect is not reducible to technical compromise. Their outcome is cognitive and social: they modify perception, confidence, expectations, and decision-making.

\subsection{Foreign information manipulation and interference}

European policy discussions increasingly use the concept of \textbf{Foreign Information Manipulation and Interference (FIMI)} to distinguish coordinated manipulative behaviour from ordinary disagreement, misinformation, or partisan communication \cite{eeas2026fourthFIMI}. The shift is important because FIMI analysis focuses less on the truth-value of an isolated message and more on patterns of behaviour, coordination, concealment, operational intent, and strategic effect.

For the purposes of this paper, FIMI is not treated as a purely geopolitical category. Rather, it provides a useful analytical orientation: manipulation should be assessed at the level of behaviours, infrastructures, and campaign dynamics. This orientation is also applicable to non-state actors, proxy networks, ideological communities, private contractors, and economically motivated manipulation.

\subsection{The operational gap}

A practical gap remains. On the one hand, incident-level analysis is too narrow. A single post, video, account, or hashtag rarely contains enough information to determine whether a manipulation campaign exists. On the other hand, strategic attribution is often too demanding. Establishing the ultimate actor behind an operation may require intelligence, legal evidence, or inter-institutional assessment that is unavailable during early response.

This creates a need for an intermediate object of analysis. Analysts must be able to say: these incidents, behaviours, assets, and narratives appear to form a coherent manipulation process, even if the final sponsor is not yet known. This is the role assigned here to Information Manipulation Sets (\textbf{IMI}s).

\subsection{Why human-centred analysis is necessary}

The operational gap is not merely technical. It is also institutional and normative. Social cyber attacks often occur in spaces where legitimate disagreement, emotional mobilisation, satire, activism, journalism, and manipulation overlap. A system that treats all hostile narratives as threats will over-securitise public debate. A system that waits for complete proof of hostile sponsorship may fail to respond to manipulation while it is operationally effective.

Human-centred analysis is therefore required because the central problem is decision-making under uncertainty \cite{nist2023humanCentered,pollini2021humanFactors}. Machine-assisted tools can identify patterns, but analysts must interpret those patterns in context, assess alternative explanations, and recommend proportionate interventions.

\section{Information Manipulation Sets as an Operational Unit}

\subsection{Intermediate unit}

AI-enabled social cyber attacks create a level-of-analysis problem. At the lowest level, analysts observe artefacts: posts, comments, videos, domains, accounts, hashtags, generated images, or bursts of interaction. At the highest level, institutions may seek strategic attribution: who organised, financed, authorised, or benefited from the operation.

Neither level is sufficient during the operational phase. Incident-level analysis is too fragmented, because individual artefacts are often ambiguous or weakly informative. Attribution-first analysis is too demanding, because responsibility may remain uncertain while the campaign is already shaping perception and behaviour. An intermediate unit is therefore required: broad enough to capture coordination across heterogeneous traces, but cautious enough not to collapse suspicion into attribution.

\subsection{Definition and relation to VIGINUM/EEAS}

An \textbf{Information Manipulation Set} is a structured collection of observable elements that appear to participate in the same manipulation process. These elements may include contents, accounts, channels, domains, media assets, hashtags, narratives, posting behaviours, amplification patterns, timing regularities, cross-platform migrations, and infrastructural traces. The IMS is not defined by a single message, platform, or definitive knowledge of the ultimate actor. It is a \textbf{hypothesis-bearing analytical object}: it organises evidence, uncertainty, alternative explanations, and mitigation options, but does not by itself prove responsibility.

This paper does not introduce the IMS concept ex nihilo. It builds on the VIGINUM/EEAS operational notion of IMS in counter-FIMI analysis \cite{viginum2026ims,eeas2026fourthFIMI}. In that context, IMS terminology supports the grouping of manipulative behaviours, tools, techniques, procedures, and resources that may be associated with the same actor or group of actors even when definitive attribution is unavailable \cite{viginum2026ims}.

The present paper repositions IMS within a SOCMINT risk-mitigation framework for AI-mediated information manipulation. Conceptually, IMS plays a role analogous to intrusion sets in Cyber Threat Intelligence (\textbf{CTI}): it groups heterogeneous traces into a provisional operational object without collapsing that object into final attribution \cite{oasisSTIXIntro,viginum2026ims}. The extension is threefold: IMS construction is connected to AI-enabled semantic variation, synthetic or semi-synthetic amplification, and human-centred mitigation rather than attribution shortcuts \cite{goldstein2023generative}.

\subsection{Boundaries and levels of analysis}

An IMS is not a single false claim, a hashtag, a community, an actor, or proof of illegitimacy. It may contain any of these elements, but it is defined by their coherence as a manipulation process.

\begin{center}
\scriptsize
\setlength{\tabcolsep}{2pt}
\renewcommand{\arraystretch}{1.05}
\begin{tabular}{@{}p{0.16\textwidth}p{0.23\textwidth}p{0.31\textwidth}p{0.22\textwidth}@{}}
\toprule
\textbf{Level} & \textbf{Unit} & \textbf{Main question} & \textbf{Typical output} \\
\midrule
Incident & Post, account, URL, video, hashtag, domain, burst & What happened? & Event description \\
Pattern & Repetition, timing, semantic convergence, network structure & Are traces related? & Cluster or anomaly \\
IMS & Behaviours, assets, narratives, infrastructures, TTPs & Do traces form a manipulation process? & Provisional operational object \\
Attribution & Operator, sponsor, contractor, state, organisation & Who is responsible? & Strategic, political, or legal claim \\
\bottomrule
\end{tabular}
\end{center}

The IMS is the level at which mitigation can begin before attribution is complete.

\subsection{Coherence, confidence, and severity}

An IMS may contain narrative, content, actor, network, temporal, infrastructural, platform, behavioural, cognitive, and AI-mediated components. Coherence can be semantic, temporal, network-based, infrastructural, behavioural, target-related, or cognitive. The more dimensions converge, the stronger the IMS hypothesis. Yet coherence is not proof: organic mobilisation can also produce similarity. The framework therefore requires alternative explanations and explicit confidence levels.

IMS analysis should separate \textbf{confidence} from \textbf{severity}. Confidence concerns the strength of the manipulation hypothesis. Severity concerns potential harm if the hypothesis is correct. Low confidence with high severity may justify urgent monitoring; high confidence with low severity may require only documentation. This distinction prevents treating weak evidence as sufficient for strong action, while also avoiding paralysis when uncertain signals target critical institutions.

Attribution should be graduated: tactical attribution identifies traces; operational attribution groups them into an IMS; strategic attribution connects the IMS to operators or sponsors; political or legal attribution assesses responsibility and response. The analytical advantage of IMS is not earlier certainty. It is structured uncertainty.

\section{AI as an Accelerator of Manipulation}

AI does not invent disinformation, propaganda, social engineering, or reputational attack; it changes their economics \cite{goldstein2023generative,bontcheva2023generativeAI}. It lowers the cost of producing plausible content, adapting messages to audiences, translating and localising narratives, simulating personas, testing frames, and flooding the information space.

Generative AI makes it easier to produce many versions of a message that are semantically aligned but lexically different \cite{goldstein2023generative}. This weakens detection based on exact duplication. A campaign can maintain narrative convergence while avoiding obvious textual repetition. AI-assisted personas can also create artificial consensus by commenting, asking leading questions, reinforcing group identity, or simulating ordinary users. The goal is not only to publish content, but to create a social environment in which an interpretation appears spontaneous and widespread.

AI also supports localisation and flooding. The same strategic frame can be adapted to different languages, communities, and cultural contexts. Volume and variability can overwhelm verification, moderation, and attention. In such cases, the objective is not persuasion alone, but degradation of the informational environment.

A further attack surface appears when users delegate search, synthesis, and preliminary judgement to AI-mediated systems. Manipulation may target not only human audiences, but also the sources from which automated systems retrieve, summarise, or contextualise answers \cite{bontcheva2023generativeAI}. The risk is not simply that a model generates a false statement. The risk is that coordinated contamination of the information environment changes what later appears to be a plausible synthesis.

For IMS-based SOCMINT, the relevant question is therefore not whether a message was generated by AI, but whether AI-enabled activity participates in a coherent manipulation process: semantic convergence, coordinated timing, infrastructure reuse, shared targets, cross-platform propagation, recurrent emotional framing, unnatural interaction patterns, and persistence across AI-mediated information sources.

\section{A SOCMINT Pipeline for IMS-Based Risk Mitigation}

The framework organises SOCMINT-based risk mitigation into an iterative pipeline:

\textbf{Signal Detection $\rightarrow$ Diagnostic Triage $\rightarrow$ IMS Hypothesis Construction $\rightarrow$ Confidence/Severity Assessment $\rightarrow$ Mitigation Selection $\rightarrow$ Feedback and Update}

The pipeline is not fully automated. It follows the human-centred cybersecurity view that technical tools should support, rather than replace, situated human judgement \cite{nist2023humanCentered,pollini2021humanFactors}. Its central output is not a content label, account score, or attribution claim, but an auditable IMS hypothesis: what may belong together, how confident analysts are, how severe the harm may be, and which responses are proportionate.

\textbf{Signal detection} identifies traces that may indicate abnormal, coordinated, or adversarial activity: posting bursts, semantic convergence, anomalous amplification, cross-platform repetition, new account clusters, reputational targeting, infrastructure reuse, or persistent frames likely to be retrieved by AI-mediated systems. Detection should capture early signals without treating ordinary controversy as manipulation.

\textbf{Diagnostic triage} asks whether signals deserve IMS construction. Analysts compare possible explanations: ordinary debate, legitimate activism, media reaction, platform-driven amplification, or coordinated manipulation. The goal is to avoid both over-detection and under-detection.

\textbf{IMS hypothesis construction} defines provisional boundaries: included contents, accounts, channels, domains, narratives, temporal patterns, infrastructures, targets, cognitive vulnerabilities, and alternative explanations. The hypothesis is revisable: new evidence can expand, split, merge, weaken, or discard it.

\textbf{Confidence/severity assessment} separates evidence strength from potential harm. Severity depends on target vulnerability, audience reach, narrative plausibility, emotional intensity, institutional sensitivity, diffusion speed, cross-platform persistence, reputational damage, likely behavioural effects, escalation potential, and persistence in search or AI-mediated synthesis environments.

\textbf{Mitigation selection} translates confidence and severity into proportionate action. Low-friction responses include monitoring, briefing, documentation, clarification, and communication preparation. Medium-friction responses include prebunking, targeted clarification, platform notification, friction mechanisms, and exposure of coordination when confidence is sufficient. High-friction responses include formal attribution, legal escalation, infrastructure disruption, sanctions, or public denunciation.

The pipeline is iterative, in line with exercise-based and incident-response approaches in cybersecurity preparedness \cite{nist2006sp80084,enisa2026exerciseMethodology}. After mitigation, analysts evaluate whether the response reduced risk, amplified the narrative, affected legitimate debate, triggered adaptation, or changed the confidence/severity assessment. The pipeline therefore turns heterogeneous SOCMINT outputs into an auditable decision process.

\section{Indicators and Observables}

\subsection{Purpose}

The indicator framework links SOCMINT observables to IMS-based risk assessment. The indicators are not designed to produce automatic attribution. They support the construction, revision, and evaluation of IMS hypotheses, drawing more on structured threat-intelligence observables and disinformation tactics, techniques, and procedures (\textbf{TTP}) modelling than on content classification alone \cite{disarm2023edmo,disarmFramework,oasisSTIXIntro,nist2016sp800150}.

The central analytical questions are: which traces may belong together, why they may form a coherent manipulation process, and what confidence and severity levels are justified.

\subsection{Indicator families}

\begin{center}
\scriptsize
\setlength{\tabcolsep}{2pt}
\renewcommand{\arraystretch}{1.05}
\begin{tabular}{@{}p{0.22\textwidth}p{0.38\textwidth}p{0.32\textwidth}@{}}
\toprule
\textbf{Family} & \textbf{What it captures} & \textbf{Examples} \\
\midrule
Content and semantic & Recurrent claims, frames, slogans, images, memes, paraphrases & lexical convergence, topic coherence, frame recurrence \\
Network & Accounts, pages, channels, communities, brokers, amplifiers & centrality, modularity, bridge nodes, repeated amplifiers \\
Temporal & Timing, bursts, synchronisation, campaign phases & simultaneous reposting, repeated waves, crisis-triggered activity \\
Cross-platform & Migration across platforms and formats & fringe-to-mainstream movement, format adaptation, platform hopping \\
Infrastructure & Domains, URLs, media assets, account creation patterns & reused domains, shared links, templates, metadata, link shorteners \\
Behavioural & Posting style and interaction patterns & comment flooding, brigading, artificial consensus, coordinated reporting \\
Target and cognitive risk & Targeted entities and expected human effects & distrust, fear, anger, humiliation, polarisation, reputational erosion \\
AI-mediated environment & Persistence and retrievability of frames & repeated paraphrases, cross-domain persistence, citation loops \\
\bottomrule
\end{tabular}
\end{center}

These families draw on network-science and information-diffusion literature \cite{bessi2015science,delvicario2016spreading,cinelli2020covid,cinelli2021echo}. They also reflect FIMI and Cyber Threat Intelligence (CTI) practice, where heterogeneous observables are organised into structured threat objects rather than treated as isolated contents \cite{viginum2026ims,oasisSTIXIntro,nist2016sp800150}. The families are not independent modules: analysts should look for convergence across them.

\subsection{Confidence and severity assessment}

Confidence in an IMS hypothesis should be based on convergence across coherence dimensions, not on a single indicator. Low confidence means weak signals and strong organic explanations. Medium confidence means several indicators converge but uncertainty remains. High confidence means strong coordination signals and limited plausible organic explanation.

Severity assesses potential harm if the IMS hypothesis is correct. Relevant dimensions include target sensitivity, audience reach, narrative plausibility, emotional intensity, likely behavioural effects, persistence, and AI-mediated retrievability. Severity should not be inferred mechanically from confidence: a low-confidence signal may require urgent monitoring if it targets a critical institution, while a high-confidence IMS may require limited action if expected harm is low.

\subsection{From indicators to decisions}

Indicators should not be interpreted mechanically. High posting frequency may reflect legitimate mobilisation; shared vocabulary may reflect common identity; cross-platform diffusion may reflect ordinary media attention; and emotional intensity may reflect justified concern. The analytical task is to assess whether several indicators converge into a coherent manipulation process and whether alternative explanations remain plausible.

A human-centred SOCMINT system should therefore ask: what has been observed; which elements appear connected; which coherence dimensions are present; what alternatives remain; what confidence and severity levels are justified; which mitigation options are proportionate; and what risks are created by responding or not responding. The framework operationalises the core claim of the paper: AI-enabled social cyber attacks are better analysed by asking whether heterogeneous traces cohere into a manipulation process, not merely whether a single message is false or AI-generated.

\section{Illustrative Scenario: Why IMS Matters}

Consider the following stylised scenario: a public institution announces a technically complex policy measure during a period of social tension. The official note contains no false statement, but one passage is ambiguous. Within hours, the institution becomes the target of an online reputational attack. Some criticism is genuine; other signals suggest that a manipulation process may be exploiting the ambiguity.

The scenario is not presented as an empirical case study. It is used to illustrate how IMS-based SOCMINT organises analysis and mitigation, and why incident-level and attribution-first approaches may miss the structure of the operation.

\begin{center}
\scriptsize
\setlength{\tabcolsep}{2pt}
\renewcommand{\arraystretch}{1.05}
\begin{tabular}{@{}p{0.22\textwidth}p{0.38\textwidth}p{0.32\textwidth}@{}}
\toprule
\textbf{Phase} & \textbf{Description} & \textbf{Key IMS signal} \\
\midrule
Trigger and ambiguity & A sentence is isolated and framed as concealment. & Real ambiguity becomes a cognitive vulnerability. \\
Semantic variation & Accounts repeat the same accusation in different wording. & Lexical variation masks semantic convergence. \\
Synthetic emotionalisation & AI-generated images and videos dramatise harm. & Affective artefacts reinforce the same frame. \\
Cross-platform migration & The narrative moves from small channels to major platforms. & Bridge accounts connect communities. \\
Apparent confirmation & News-like domains repeat and cite the allegation. & A loop creates apparent independent evidence. \\
Pressure for overreaction & Posts frame silence or denial as guilt. & The campaign targets institutional decision-making. \\
\bottomrule
\end{tabular}
\end{center}

No single incident is decisive. The posts are not identical, the accounts do not all interact directly, the domains do not openly share ownership, and some amplifiers are authentic citizens. Incident-level analysis fragments the campaign into weak signals. Attribution-first analysis also fails during the operational phase because no sponsor can yet be identified. The IMS approach captures what both perspectives miss: the recurrence of the same operational sequence across heterogeneous traces.

Applied to the pipeline, detection identifies bursts, semantic convergence, synthetic visuals, new accounts, cross-platform migration, bridge accounts, and news-like domains. Triage compares organic criticism, media dynamics, platform amplification, and coordinated manipulation. IMS construction groups the repeated betrayal/concealment frame, varied messages, generated visuals, bridge accounts, similar domains, and pressure tactics. Confidence may be medium, because several coherence dimensions converge but organic criticism remains plausible; severity may be high, because the target is sensitive and overreaction risk is significant.

A proportionate response would monitor the IMS, brief decision-makers, clarify the ambiguity, avoid repeating the hostile frame, notify platforms only about clearly inauthentic behaviour, and preserve evidence. The scenario illustrates the role of IMS as an intermediate analytical unit: the campaign is visible not in any single artefact, but in the repeated organisation of heterogeneous traces.

\section{Human-Centred Mitigation and Democratic Safeguards}

AI-assisted systems can detect anomalies, cluster narratives, identify suspicious timing, and summarise large data streams. Mitigation, however, cannot be delegated entirely to automated systems. The relevant decisions involve context, proportionality, uncertainty, legal constraints, institutional credibility, and democratic legitimacy.

The central safeguard is the distinction between legitimate dissent and coordinated manipulation. A democratic society must tolerate criticism, anger, satire, polarisation, and even false or poorly informed opinions. These phenomena are not automatically social cyber attacks. Additional features are required: coordination, concealment, artificial amplification, deceptive infrastructure, strategic targeting, or other behaviours indicating manipulation of the information environment. The framework therefore focuses on patterns of behaviour, not ideological content alone.

Mitigation should be proportionate to both confidence and severity. A low-confidence IMS should normally trigger monitoring, not public accusation. A high-confidence but low-severity IMS may require documentation rather than intervention. A high-confidence, high-severity IMS may justify public exposure, platform notification, or institutional escalation. Proportionality also requires considering second-order effects: a response can amplify a narrative, legitimise a marginal campaign, or appear to suppress dissent.

Analysts remain responsible for interpreting context, assessing alternatives, distinguishing organic mobilisation from manipulation, assigning confidence levels, recommending interventions, documenting uncertainty, and preventing over-securitisation. Human judgement is not an obstacle to scalability; it is the layer that prevents scalable systems from producing scalable errors.

Cognitive security also depends on institutional credibility. A technically correct response may fail if the responding institution is not trusted. Conversely, premature or exaggerated accusations can damage credibility and make future warnings less effective. For this reason, mitigation cannot be separated from strategic communication (\textbf{StratCom}) \cite{eeas2024secondFIMI,eeas2026fourthFIMI}.

Where possible, IMS reports should distinguish observation, inference, confidence, and recommended action. A human-centred design principle follows: machine-assisted systems should detect, organise, and explain signals; human analysts should interpret, contextualise, and decide; institutions should act proportionately and remain accountable.

\section{Evaluation Strategy: Tabletop Assessment}

Because the contribution is a conceptual-operational framework rather than a predictive model, evaluation should not be reduced to classification accuracy. The relevant question is whether the framework improves decision quality under uncertainty: earlier recognition of coherent manipulation, better distinction between dissent and manipulation, calibrated confidence/severity assessment, proportionate mitigation, and auditable reasoning.

A practical evaluation can use a tabletop exercise, a discussion-based method used to test roles, coordination, and decision-making in cybersecurity and incident response \cite{nist2006sp80084,enisa2026exerciseMethodology}. Participants receive a simulated AI-enabled operation through staged injects: ambiguous trigger, early criticism, semantic variation, synthetic emotionalisation, cross-platform migration, apparent confirmation, pressure for reaction, mitigation decision, and after-action update.

At each stage, participants identify signals, decide whether IMS construction is justified, define boundaries, list alternatives, assign confidence and severity, recommend mitigation, and state what evidence would change their assessment. The exercise can compare three conditions: incident-level analysis, attribution-first analysis, and IMS-based analysis. The comparison follows scenario-based evaluation: staged injects are used to assess decisions, updates, and documentation over time.

Evaluation should consider process quality, reasoning quality, calibration, mitigation proportionality, and safeguards. Examples include time to IMS hypothesis, update quality, analyst workload, alternative explanations, rejected hypotheses, audit trail, confidence/severity calibration, second-order risk awareness, dissent/manipulation distinction, and premature attribution count. The expected advantage of IMS is not earlier certainty. It is better structured uncertainty.

\subsection{Boundary check for evaluation design}

The tabletop protocol is complemented by a boundary check. This check is not empirical validation; it tests whether the IMS lens preserves distinctions essential to human-centred cybersecurity. The framework should help analysts distinguish adversarial manipulation from organic mobilisation, institutional communication failure, market panic, or emergent collective behaviour, rather than classify every rapid mobilisation, reputational crisis, or high-visibility controversy as coordinated manipulation.

\begin{center}
\scriptsize
\textbf{Table 1.} Boundary check for IMS-based SOCMINT.\par\vspace{2pt}
\begin{tabular}{p{.23\linewidth}p{.35\linewidth}p{.32\linewidth}}
\toprule
Case type & IMS-relevant question & Expected analytical use \\
\midrule
Election-related manipulation & Do heterogeneous traces cohere before final attribution is possible? & Tests IMS as an intermediate unit between incident detection and attribution. \\
Institutional reputational pressure & Do impersonation, synthetic content, cross-platform circulation, and targeting form an operational sequence? & Tests whether IMS connects artefacts, reputational pressure, and mitigation choices. \\
Critical-infrastructure narratives & Does narrative exploitation of physical or cyber incidents form a distinct IMS component, or does it reflect independent organic reaction? & Tests whether IMS prevents over-attribution across domains. \\
Financial panic or trust collapse & Is rapid diffusion evidence of manipulation, or organic networked behaviour? & Serves as a negative control: tests whether the framework avoids over-securitising organic financial stress. \\
Community-driven mobilisation & Is coordination public and emergent rather than covert and deceptive? & Tests the distinction between collective action and manipulation. \\
\bottomrule
\end{tabular}
\end{center}

The table defines boundary conditions rather than case-study results. The supplementary appendix provides optional scenario material and retrospective case notes that implementers may use to instantiate these conditions in tabletop exercises. It is not required for understanding or evaluating the framework.

The boundary check supports evaluation by clarifying what should and should not count as an IMS. Its purpose is not to prove detection performance, but to discipline attribution and prevent conceptual overextension.

\section{Discussion and Limitations}

The first limitation is false positives. Online mobilisation can be rapid, emotional, repetitive, and polarised without being coordinated manipulation. Communities may independently converge on similar language because they share identity, sources, or political interpretation. IMS construction must therefore remain probabilistic and transparent.

The second limitation is false negatives. Sophisticated actors may avoid obvious synchronisation, diversify infrastructures, use authentic users, exploit organic communities, and generate semantically varied content. AI can make manipulation appear less repetitive and more culturally adapted.

Third, attribution remains politically and legally sensitive. IMS analysis can support operational attribution, but strategic or political attribution requires additional evidence and institutional procedures. The paper therefore separates IMS construction from final responsibility claims.

Fourth, SOCMINT depends on data access. Platform restrictions, API limits, deleted content, encrypted channels, private groups, and algorithmic opacity constrain what analysts can observe. The framework must operate under incomplete observability.

Fifth, adversaries adapt. Once indicators become known, they can vary timing, diversify language, distribute assets, hide coordination, use unwitting amplifiers, or contaminate legitimate debate. The pipeline must therefore remain iterative.

Finally, SOCMINT can create privacy and civil-liberties risks. The framework requires data minimisation, purpose limitation, access control, and oversight. Human-centred cybersecurity includes protection from both adversarial manipulation and disproportionate surveillance.

The paper provides an evaluation strategy but not completed exercise results. Future work should implement the protocol with analysts, security practitioners, communication officers, and domain experts to test confidence calibration, response proportionality, and the distinction between dissent and manipulation.

\section{Conclusion}

AI-enabled social cyber attacks exploit the intersection of digital infrastructures, platform dynamics, social behaviour, and cognitive vulnerability. Their security relevance cannot be understood by focusing only on isolated false contents or on the detection of AI-generated text. The relevant object is often a manipulation process: a coordinated set of behaviours, narratives, assets, infrastructures, temporal patterns, and amplification mechanisms.

This paper proposed a SOCMINT framework based on Information Manipulation Sets. The IMS is an intermediate operational unit between incident-level detection and full strategic attribution. It allows analysts to group observable traces into a coherent hypothesis while preserving uncertainty and avoiding premature attribution.

The proposed pipeline connects computational support with analyst judgement. It treats AI as an accelerator of manipulation, but does not reduce the problem to AI detection. It treats SOCMINT indicators as evidence for structured reasoning, not as automatic proof. It treats mitigation as human-centred and institutionally accountable.

The paper also argued that framework evaluation should focus on decision quality under uncertainty. A tabletop exercise can test whether IMS-based analysis improves confidence assignment, alternative-explanation tracking, the distinction between dissent and manipulation, and mitigation proportionality.

The main implication is that cognitive security must be designed as a socio-technical capability. It requires machine learning, NLP, network analysis, platform intelligence, strategic communication, institutional credibility, and democratic safeguards. The goal is not to suppress disagreement, but to identify and mitigate coordinated manipulation that distorts the conditions of public judgement and collective decision-making.

\bibliographystyle{splncs04}
\bibliography{references}
\end{document}